\documentclass[aps,prd,showpacs,twocolumn,floatfix,nofootinbib]{revtex4-1}
\usepackage[dvipdfmx]{graphicx}
\usepackage{amsmath,amssymb,siunitx}
\usepackage{color}

\begin{document}
\title{Detectability of thermal neutrinos from binary neutron-star
mergers and implication to neutrino physics}
\author{
Koutarou Kyutoku$^{1,2,3,4}$ and Kazumi Kashiyama$^5$
}
\affiliation{
$^1$Theory Center, Institute of Particle and Nuclear Studies, KEK,
Tsukuba 305-0801, Japan\\
$^2$Department of Particle and Nuclear Physics, the Graduate University
for Advanced Studies (Sokendai), Tsukuba 305-0801, Japan\\
$^3$Interdisciplinary Theoretical Science (iTHES) Research Group, RIKEN,
Wako, Saitama 351-0198, Japan\\
$^4$Center for Gravitational Physics, Yukawa Institute for Theoretical
Physics, Kyoto University, Kyoto 606-8502, Japan\\
606-8502, Japan\\
$^5$Department of Physics, the University of Tokyo, Bunkyo, Tokyo
113-0033, Japan\\
}
\date{\today}

\begin{abstract}
 We propose a long-term strategy for detecting thermal neutrinos from
 the remnant of binary neutron-star mergers with a future M-ton
 water-Cherenkov detector such as Hyper-Kamiokande. Monitoring $\gtrsim
 \num{2500}$ mergers within $\lesssim \SI{200}{Mpc}$, we may be able to
 detect a single neutrino with a human time-scale operation of $\approx
 \SI{80}{Mt.years}$ for the merger rate of \SI{1}{Mpc^{-3}.Myr^{-1}},
 which is slightly lower than the median value derived by the LIGO-Virgo
 Collaboration with GW170817. Although the number of neutrino events is
 minimal, contamination from other sources of neutrinos can be reduced
 efficiently to $\approx 0.03$ by analyzing only $\approx
 \SI{1}{\second}$ after each merger identified with gravitational-wave
 detectors if gadolinium is dissolved in the water. The contamination
 may be reduced further to $\approx 0.01$ if we allow the increase of
 waiting time by a factor of $\approx 1.7$. The detection of even a
 single neutrino can pin down the energy scale of thermal neutrino
 emission from binary neutron-star mergers and could strongly support or
 disfavor formation of remnant massive neutron stars. Because the
 dispersion relation of gravitational waves is now securely constrained
 to that of massless particles with a corresponding limit on the
 graviton mass of $\lesssim \SI{e-22}{\eV} / c^2$ by binary black-hole
 mergers, the time delay of a neutrino from gravitational waves can be
 used to put an upper limit of $\lesssim O(10)\,\si{\milli\eV} / c^2$ on
 the absolute neutrino mass in the \emph{lightest} eigenstate. Large
 neutrino detectors will enhance the detectability, and, in particular,
 \SI{5}{\mega\tonne} Deep-TITAND and \SI{10}{\mega\tonne} MICA planned
 in the future will allow us to detect thermal neutrinos every $\approx
 16$ and \SI{8}{years}, respectively, increasing the significance.
\end{abstract}
\pacs{04.30.Tv, 14.60.Pq, 95.85.Ry}

\maketitle

\section{Introduction} \label{sec:intro}

The discovery of GW170817 marked an opening of multimessenger astronomy
with binary neutron-star mergers
\cite{ligovirgo2017-3,ligovirgoem2017}. Gravitational waves driving the
mergers are important targets for ground-based detectors such as
Advanced LIGO and Advanced Virgo and enable us to study the equation of
state of supranuclear density matter \cite{ligovirgo2017-3}. Differently
from binary black holes, binary neutron stars can also become bright in
electromagnetic channels \cite{metzger_berger2012,piran_nr2013}. The
merger remnants are the prime candidate for the central engine of
short-hard gamma-ray bursts
\cite{paczynski1986,goodman1986,eichler_lps1989,narayan_pp1992} as has
already been suggested by association of GRB 170817A with GW170817
\cite{ligovirgogamma2017} (but see
Refs.~\cite{mooley_etal2018,ruan_nhke2018}). Binary neutron stars should
also eject neutron-rich material during the merger and postmerger
phases, and this material will be synthesized to heavy neutron-rich
nuclei, namely the \textit{r}-process elements
\cite{lattimer_schramm1974,symbalisty_schramm1982}. This scenario is
supported by detections of electromagnetic counterparts consistent with
the macronova/kilonova, an optical-infrared transient powered by decay
of the \textit{r}-process elements \cite{li_paczynski1998}, after
GW170817 (see, e,g., Ref.~\cite{ligovirgoem2017}).

Signals from binary neutron-star mergers are not limited to
gravitational and electromagnetic radiation. Because the violent merger
heats up the high-density material, thermal neutrinos with $\gtrsim
\SI{10}{\mega\eV}$ should also be emitted from the remnant of binary
neutron-star mergers
\cite{ruffert_jts1997,rosswog_liebendorfer2003,sekiguchi_kks2011}.

While direct detections of thermal neutrinos could be an important step
toward understanding the realistic merger process as well as their
impact on gamma-ray bursts \cite{just_ojbs2016,fujibayashi_sks2017} and
\textit{r}-process nucleosynthesis
\cite{wanajo_snkks2014,goriely_bjpj2015}, they are quite challenging. On
one hand, as we will see later, the detection is hopeless for a single
merger at a distance $\gtrsim \SI{100}{Mpc}$ where the mergers are
expected to occur more than once a year
\cite{ligovirgo2010,dominik_bombfhbp2015}. On the other hand, ``the
diffuse neutron-star-merger neutrino background,'' i.e., superposition
of neutrinos from all the binary neutron-star mergers throughout the
Universe, is inevitably hidden by the diffuse supernova neutrino
background also known as supernova relic neutrinos (see
Ref.~\cite{beacom2010} for reviews), because the rate of supernovae must
be higher by a few orders of magnitude than that of binary neutron-star
mergers. However, we would like to stress again that detecting thermal
neutrinos will be important to understand binary neutron-star mergers
accurately, as theoretical models of supernova explosions are
qualitatively confirmed with detections of neutrinos from SN 1987A
\cite{hirata_etal1987,bionta_etal1987}.

The chance of detections lies in the gap between these two standard
ideas (see also Refs.~\cite{ando_by2005,kistler_yabs2011,boser_kssv2015}
for a third idea on detecting supernova neutrinos). In this paper, we
propose a long-term strategy to detect thermal neutrinos from the
remnant of binary neutron-star mergers by monitoring many mergers
identified by gravitational-wave detectors. Stacking multiple mergers is
necessary except for serendipitous nearby mergers, but careless analysis
will easily bury \si{\mega\eV} neutrinos from binary neutron-star
mergers in those from other sources such as the diffuse supernova
neutrino background and atmospheric neutrinos including invisible muons
(see Ref.~\cite{kimura_mmk2017} for stacking of high-energy
neutrinos). A remarkable point is that gravitational-wave observations
may determine the time of merger to an accuracy of $\approx
\SI{1}{\milli\second}$. Indeed, this precision has already been realized
for observed binary black-hole mergers
\cite{ligovirgo2016-5}.\footnote{Because the gravitational-wave
frequency at mergers of binary neutron stars is likely to be higher than
the values accessible by current detectors, we might have to rely on
theoretical models to determine the time of merger in realistic
situations. The time of merger can vary by up to a few
\si{\milli\second} depending on the equation of state of neutron-star
matter. We expect that the situation will improve by constraining the
equation of state using gravitational-wave observations themselves.} By
analyzing the data of neutrino detectors only during $\approx
\SI{1}{\second}$ from each merger, which is much shorter than
\SI{1000}{\second} adopted in current counterpart searches with
MeV-neutrino detectors \cite{kamland2016,sk2016}, we can substantially
reduce contamination from other sources of neutrinos. A similar idea has
been proposed for detecting gravitational waves informed by gamma-ray
bursts \cite{kochanek_piran1993}.

If the arrival time of such neutrinos with respect to the time of merger
identified by gravitational waves is successfully determined with modest
delay of 0.1--\SI{1}{\second}, we may be able to put an upper limit on
the absolute neutrino mass in the \emph{lightest} eigenstate
\cite{nishizawa_nakamura2014,langaeble_ms2016} (see also
Ref.~\cite{baker_trodden2017}). Because neutrinos have finite masses of
$\lesssim \SI{0.1}{\eV} / c^2$, their travel speed is necessarily slower
than the speed of light $c$, where the precise value depends on the mass
eigenstate and the energy. Importantly, detections of gravitational
waves from binary black-hole mergers have successfully shown that the
dispersion relation of gravitational waves is accurately described by
that of massless particles with a corresponding limit on the gravitons
mass of $\lesssim \SI{e-22}{\eV} / c^2$
\cite{ligovirgo2016-5,ligovirgo2017}, practically negligible. Thus, the
difference of arrival times between these two messengers, or relative
time of flight, will allow us to infer the mass of neutrinos. Although
chances are not necessarily large, the range of accessible masses seems
worth of serious consideration.

This paper is organized as follows. First, we summarize current
understanding of neutrino emission from binary neutron-star mergers in
Sec.~\ref{sec:model}. Next, we describe our strategy to detect thermal
neutrinos from binary neutron-star mergers in Sec.~\ref{sec:det}, and
the expected level of contamination is examined in
Sec.~\ref{sec:contami}. Implication of detecting thermal neutrinos is
discussed in Sec.~\ref{sec:phys}. Section \ref{sec:summary} is devoted
to a summary. While we focus only on binary neutron stars in this work
due to their expected dominance, it is straightforward to enhance our
discussion to include black hole--neutron star binaries.

\section{Thermal neutrino from binary neutron-star mergers}
\label{sec:model}

We first summarize characteristics of thermal neutrinos emitted from the
remnant of binary neutron-star mergers. In this work, we focus on the
case that the lifetimes of remnant neutron stars are longer than $\sim
\SI{1}{\second}$ and thus substantial neutrino emission can be
expected. This is not for optimistic simplification, but the scenario
that we would like to verify or reject by detecting thermal
neutrinos. The lifetime of remnant massive neutron stars depends on
various details and can in fact be very short
\cite{sekiguchi_kks2011,hotokezaka_kkmsst2013}. Fortunately, the prompt
collapse to a black hole is not very likely in light of the maximum mass
of spherical neutron stars exceeding $\approx 2 M_\odot$
\cite{hotokezaka_kosk2011}.

Various numerical simulations have shown that the remnant massive
neutron stars are heated up to several tens of \si{\mega\eV} at the
collision unless the merger results in a prompt collapse
\cite{ruffert_jts1997,rosswog_liebendorfer2003,sekiguchi_kks2011}. Then,
thermally-produced electron-positron pairs are captured on nucleons to
emit a copious amount of electron neutrinos $\nu_e$ and antineutrinos
$\bar{\nu}_e$ with the rise time of $\lesssim \SI{10}{\milli\second}$
from the merger. In the case of binary neutron-star mergers,
$\bar{\nu}_e$ is brighter than $\nu_e$ due to the neutron richness. The
peak luminosity of electron antineutrinos reaches
$1$--\SI{3e53}{erg.s^{-1}} with the typical energy 10--\SI{30}{\mega\eV}
depending on binary parameters and unknown equations of state for
supranuclear-density matter
\cite{sekiguchi_kks2015,palenzuela_lnlcoa2015,foucart_hdoorklps2016,sekiguchi_kkst2016}. Pair
processes such as the electron-positron annihilation also emit muon and
tau (anti)neutrinos. We note that neutrino oscillations in the source
region such as the Mikheyev-Smirnov-Wolfenstein effect, bipolar
oscillations, and matter-neutrino resonance begin to be explored only
recently for remnant massive neutron stars
\cite{zhu_pm2016,frensel_wvp2017,chatelain_volpe2017,wu_tamborra2017}. They
could reduce effective luminosity of electron antineutrinos to some
extent.

So far, little is known about the realistic spectrum of neutrinos from
binary neutron-star mergers despite their importance for quantifying the
detectability. Monte-Carlo neutrino-transport simulations suggest that
the spectrum can be approximated by pinched Fermi-Dirac distribution for
binary neutron-star mergers as in the case of supernova explosions
\cite{richers_kofo2015}. However, it is difficult to determine the
degree of distortion at this stage.

The duration of neutrino emission is expected to be a few to ten seconds
unless it is shut down by the collapse of the remnant during the hot
phase due to angular-momentum redistribution \cite{sekiguchi_kks2011},
whereas detailed long-term calculations are not available beyond
$\approx \SI{0.5}{\second}$ \cite{fujibayashi_sks2017}. The total energy
given to neutrinos is determined by hydrodynamic interactions and will
be similar to, or moderately less than, that in supernova explosions
(see Ref.~\cite{janka2016} for reviews).

\section{Detecting thermal neutrinos} \label{sec:det}

\subsection{Expected number of neutrino events}

We aim at detecting thermally-produced neutrinos from remnant massive
neutron stars formed after binary neutron-star mergers with a
water-Cherenkov detector such as planned $\approx
\SI{0.37}{\mega\tonne}$ Hyper-Kamiokande
\cite{hk2011,hk2016}.\footnote{See also
http:\slash\slash{}lib-extopc.kek.jp\slash{}preprints\slash{}PDF\slash{}2016\slash{}1627\slash{}1627021.pdf
for a recent design of Hyper-Kamiokande.}  Water-Cherenkov detectors are
efficient at detecting electron antineutrinos via the inverse $\beta$
decay, $p + \bar{\nu}_e \to n + e^+$. The expected number of neutrino
events for a single merger is estimated by
\begin{equation}
 N_\nu = N_T \int_{t_i}^{t_f} \int_{E_\mathrm{min}}^{E_\mathrm{max}}
  \phi (E,t) \sigma (E) dE dt , \label{eq:num}
\end{equation}
where $N_T$ is the number of target protons in the detector, $E$ is the
energy of electron antineutrinos, $\phi (E,t)$ is the number flux of
electron antineutrinos per unit energy, $\sigma (E)$ is the capture
cross section of an electron antineutrino on a proton. The number of
target protons is given in terms of the (effective) mass of water $M_T$
as $N_T \approx ( M_T / m_\mathrm{p} ) \times (2/18) = \num{6.7e34} (
M_T / \SI{1}{\mega\tonne} )$ with $m_\mathrm{p}$ the proton mass. The
cross section is calculated to various levels of approximations
\cite{vogel_beacom1999,strumia_vissani2003}, and in this study we adopt
Eq.~(7) of Ref.~\cite{beacom2010},
\begin{equation}
 \sigma (E) = \SI{9.5e-42}{\square\centi\meter} \left( \frac{E -
                                                 \SI{1.3}{\mega\eV}}{\SI{10}{\mega\eV}}
                                                \right)^2 \left( 1 -
                                                \frac{7E}{m_\mathrm{p}
                                                c^2} \right)
                                                . \label{eq:cs}
\end{equation}
Note that the positrons from inverse $\beta$ decay are distributed
nearly isotropically \cite{vogel_beacom1999}. Threshold energies $\{
E_\mathrm{min}, E_\mathrm{max} \}$ should be determined to span the
range relevant to neutrinos from binary neutron-star mergers while
suppressing contamination from other sources of neutrinos. Threshold
times $\{ t_i , t_f \}$ should be determined to detect intense neutrino
emission around the peak time, while the level of contamination should
be kept as low as possible.

\begin{table*}
 \caption{Contribution to $f_E$ from different energy ranges $[
 E_\mathrm{min} : E_\mathrm{max} ]$ for a variety of the typical energy
 of neutrinos, $\langle E \rangle$. Recall $f_E$ denotes the ratio of
 the number of neutrino interactions obtained by integrating the product
 of the Fermi-Dirac distribution with typical energy $\langle E \rangle
 \approx 3.15 k_\mathrm{B}T$ and the cross section, Eq.~\eqref{eq:cs},
 to the number for monoenergetic neutrinos with $\langle E \rangle$ and
 the leading-order cross section, Eq.~\eqref{eq:cslo}. The final column,
 10--\SI{50}{\mega\eV}, is the range expected to be utilizable with
 Hyper-Kamiokande with Gd dissolution \cite{beacom_vagins2004,hk2011}.}
 \begin{tabular}{ccccccc} \hline
  $\langle E \rangle$ & 0--\SI{10}{\mega\eV} & 10--\SI{20}{\mega\eV} &
  20--\SI{30}{\mega\eV} & 30--\SI{40}{\mega\eV} & 40--\SI{50}{\mega\eV}
  & 10--\SI{50}{\mega\eV} \\
  \hline \hline
  \SI{10}{\mega\eV} & 0.16 & 0.53 & 0.20 & 0.03 & $<0.01$ & 0.77 \\
  \SI{15}{\mega\eV} & 0.04 & 0.33 & 0.33 & 0.15 & 0.05 & 0.87 \\
  \SI{20}{\mega\eV} & 0.02 & 0.17 & 0.29 & 0.22 & 0.12 & 0.80 \\
%  \SI{10}{\mega\eV} & 0.26 & 0.72 & 0.27 & 0.05 & $<0.01$ & 1.05 \\ LO
%  \SI{15}{\mega\eV} & 0.07 & 0.45 & 0.46 & 0.22 & 0.08 & 1.21 \\ LO
%  \SI{20}{\mega\eV} & 0.02 & 0.23 & 0.39 & 0.32 & 0.19 & 1.14 \\ LO
  \hline
 \end{tabular}
 \label{table:fdinteg}
\end{table*}

In this work, we adopt the Fermi-Dirac distribution with temperature $T$
and zero chemical potential, for which the average energy of neutrinos
is given by $\langle E \rangle \approx 3.15 k_\mathrm{B}T$ with
$k_\mathrm{B}$ the Boltzmann constant. Thus, by ignoring the time
dependence, the spectrum or number flux per unit energy takes the form
\begin{equation}
 \phi (E) = \frac{c}{2 \pi^2 ( \hbar c )^3} \frac{E^2}{\exp [ E / (
  k_\mathrm{B}T ) ] + 1} ,
\end{equation}
where $\hbar$ is the reduced Planck constant. The expected rate of
neutrino events is obtained by integrating Eq.~\eqref{eq:num} over a
given energy interval, and the result is usefully characterized by the
typical energy, $\langle E \rangle$, and the leading-order cross section
(called ``naive'' in Ref.~\cite{strumia_vissani2003})
\begin{equation}
 \sigma_\mathrm{LO} (E) = \SI{9.5e-42}{\square\centi\meter} \left(
                                                             \frac{E}{\SI{10}{\mega\eV}}
                                                            \right)^2 ,
 \label{eq:cslo}
\end{equation}
as
\begin{equation}
 \frac{dN_\nu}{dt} \approx f_E N_T \frac{L_\nu}{4 \pi D^2}
  \frac{\sigma_\mathrm{LO} ( \langle E \rangle )}{\langle E \rangle} ,
  \label{eq:dndt}
\end{equation}
where $L_\nu$ is the luminosity of electron antineutrinos and $D$ is the
distance to the source. A factor $f_E \sim 1$ is a number determined by
$\{ E_\mathrm{min} , E_\mathrm{max} \}$ (for the assumed spectrum), and
we show contribution from different energy ranges in Table
\ref{table:fdinteg} for various values of $\langle E \rangle$. Note that
$\sigma_\mathrm{LO} (E)$ is never used to compute the expected number of
neutrino events, and we use it only for normalizing the result keeping
the leading-order dependence on $\langle E \rangle$ transparent. This
table shows that anywhere between \SI{10}{\mega\eV} and
\SI{50}{\mega\eV} can contribute appreciably to detections. Possible
choices of $\{ E_\mathrm{min} , E_\mathrm{max} \}$ will be discussed
later in Sec.~\ref{sec:contami} along with contamination from other
sources of neutrinos.

The expected number of neutrino events is obtained by integrating
Eq.~\eqref{eq:dndt} in time, but time evolution of the luminosity is not
understood in detail, particularly on a time scale of $\gtrsim
\SI{1}{\second}$. Here, we focus on $\Delta t_\mathrm{obs} \approx
\SI{1}{\second}$ from the merger and denote the total energy of electron
antineutrinos emitted during $\Delta t_\mathrm{obs}$ as $E_{\Delta t} =
\int L_\nu dt$, which may be $\approx \SI{3e52}{erg}$ for moderately
compact remnant neutron stars \cite{fujibayashi_sks2017}.\footnote{Note
that Ref.~\cite{fujibayashi_sks2017} do not incorporate viscous heating,
and thus the neutrino luminosity is likely to be underestimated.} This
restriction in time is due partly to the lack of knowledge about
long-term evolution of the neutrino luminosity, but this is mainly
intended to reduce contamination from other sources of neutrinos in
realistic observations as described later in Sec.~\ref{sec:contami}. The
total energy of neutrinos may be increased by a factor of 2--3 if we
take $\Delta t_\mathrm{obs} \approx \SI{10}{\second}$ as the available
energy budget suggests \cite{sekiguchi_kks2011}.

By regarding the typical energy of neutrinos, $\langle E \rangle$, as an
appropriate time average, the expected number of neutrino events for a
single merger is found to be
\begin{align}
 N_\nu \approx \num{1.0e-3} & \times f_E f_\mathrm{se} f_\mathrm{osc}
 \left( \frac{M_T}{\SI{1}{\mega\tonne}} \right) \left( \frac{E_{\Delta
 t}}{\SI{3e52}{erg}} \right) \notag \\
 & \times \left( \frac{\langle E \rangle}{\SI{10}{\mega\eV}} \right)
 \left( \frac{D}{\SI{100}{Mpc}} \right)^{-2} . \label{eq:numexp}
\end{align}
A fudge factor $f_\mathrm{se} \le 1$ is the event selection efficiency
for the inverse $\beta$ decay, and we expect $f_\mathrm{se} \approx 0.9$
and 0.67 for Hyper-Kamiokande without and with gadolinium (Gd),
respectively \cite{hk2011}. Another fudge factor $f_\mathrm{osc}
\lesssim 1$ represents the effect of neutrino oscillations. As discussed
in Sec.~\ref{sec:model}, the oscillation in the source region are not
fully understood yet. Therefore, we leave incorporation of these effects
for future study and discuss only the vacuum oscillation during the
propagation. In a tri-bimaximal mixing approximation
\cite{harrison_ps2002}, the survival probability of electron
(anti)neutrinos is given by $5/9$ and the appearance probability from
muon and tau (anti)neutrinos is $2/9$ each. Taking the lower luminosity
and higher typical energy of muon and tau (anti)neutrinos from the
remnant of binary neutron-star mergers
\cite{sekiguchi_kks2011,fujibayashi_sks2017}, we expect that
$f_\mathrm{osc}$ is not very far from unity. We discuss time dilation
associated with the finite masses later in Sec.~\ref{sec:phys_mass}.

This estimate, Eq.~\eqref{eq:numexp}, clearly shows that it is hopeless
to detect thermal neutrinos from a single binary neutron-star merger
except for extremely lucky stars at $\lesssim \SI{3}{Mpc}$. This fact
has already been found in previous work
\cite{sekiguchi_kks2011,palenzuela_lnlcoa2015}, and we just confirm it
with slightly detailed calculations (see also
Ref.~\cite{caballero_ms2009,liu_zlmx2016} for relevant work on accretion
flows).

\subsection{Monitoring multiple mergers}

Even though the expected number of neutrinos from a single merger is
very low, superposition of many mergers gives us a fair chance of
detections. As a qualitative order-of-magnitude estimate, we expect to
receive a single thermal neutrino with probability $1 - (1 - N_\nu
)^{1/N_\nu} \approx 1 - e^{-1} = 63\%$ with $1/N_\nu \approx \num{1000}$
mergers at \SI{100}{Mpc}. The problem is that we will have to wait
longer than a decade to collect such a large number of mergers, so that
neutrinos from other sources completely overwhelm thermal neutrinos from
binary neutron-star mergers. However and most importantly, if we focus
only on $\Delta t_\mathrm{obs} \approx \SI{1}{\second}$ after each
merger using timing information from gravitational-wave detectors, we
can efficiently reduce the length of data from neutrino detectors to
$\approx \SI{1000}{\second}$ in total. As we discuss later in
Sec.~\ref{sec:contami}, the expected number of contamination events can
be reduced to much less than unity for this short duration. We note that
neutrino detectors do not require low-latency alerts from
gravitational-wave detectors to perform this analysis
\cite{kamland2016,sk2016}.

To assess the effectiveness of this strategy in a quantitative manner,
we need to take the spatial distribution of binary neutron-star mergers
into account. In this study, we assume the effective range
$D_\mathrm{eff}$ of a gravitational-wave detector for binary neutron
stars to be \SI{200}{Mpc}, which is approximately the design sensitivity
of Advanced LIGO \cite{ligovirgo2010}. The detectable volume is given by
$4 \pi D_\mathrm{eff}^3 /3$ by definition, and the detection rate of
mergers is given by multiplying this volume by the merger rate per unit
volume per unit time of binary neutron stars, $\mathcal{R}$, which is
derived to be ${1.54}_{-1.22}^{+3.2}\,\si{Mpc^{-3}.Myr^{-1}}$ by the
LIGO-Virgo Collaboration with GW170817 \cite{ligovirgo2017-3}.

\begin{table}
 \caption{Factors appearing in the estimate of the waiting time,
 Eq.~\eqref{eq:time}, and their origins. We also present their expected
 values.}
 \begin{tabular}{cccc} \hline
  Symbol & Origin & Expected value \\
  \hline \hline
  $f_E$ & Energy range & $\approx 0.8$ (see Table
          \ref{table:fdinteg}) \\
  $f_\mathrm{se}$ & Selection efficiency & 0.9 (no Gd) or 0.67 (Gd) \\
  $f_\mathrm{osc}$ & Neutrino oscillation & $0.5 \sim 1$ \\
  $f_\Omega$ & Antenna pattern & $0.8 \sim 1$ \\
  \hline
 \end{tabular}
 \label{table:factor}
\end{table}

The period $P$ or exposure $P M_T$ that we have to wait with monitoring
multiple mergers to detect a thermal neutrino is estimated by the
condition
\begin{equation}
 1 = P \mathcal{R} f_\Omega \int_0^{D_\mathrm{eff}} N_\nu (D) \times 4
  \pi D^2 dD , \label{eq:cond}
\end{equation}
where $N_\nu (D)$ is Eq.~\eqref{eq:numexp} regarded as a function of
$D$. Here, we neglect the cosmological redshift and evolution history
that should be insignificant at $\lesssim \SI{200}{Mpc}$. A factor
$f_\Omega$ reflects the antenna pattern of gravitational-wave detectors,
i.e., dependence of the sensitivity on the sky position and orientation
of sources. We can show that $f_\Omega \approx 0.8$ for a single
detector (see Appendix \ref{sec:antenna}) and expect this to approach
unity for a network of detectors. Here, we neglect angular dependence of
neutrino emission. It should be cautioned that this waiting time, $P$,
denotes not only the operation time of the neutrino detector but also
requires coincident operations of gravitational-wave detectors.

Solving Eq.~\eqref{eq:cond}, we finally obtain
\begin{align}
 P M_T & = \SI{80}{Mt.years} \notag \\
 & \times \left( \frac{f_\mathrm{all}}{0.5} \right)^{-1} \left(
 \frac{E_{\Delta t}}{\SI{3e52}{erg}} \right)^{-1} \left( \frac{\langle E
 \rangle}{\SI{10}{\mega\eV}} \right)^{-1} \notag \\
 & \times \left( \frac{D_\mathrm{eff}}{\SI{200}{Mpc}} \right)^{-1}
 \left( \frac{\mathcal{R}}{\SI{1}{Mpc^{-3}.Myr^{-1}}} \right)^{-1} ,
 \label{eq:time}
\end{align}
where $f_\mathrm{all} \equiv f_E f_\mathrm{se} f_\mathrm{osc} f_\Omega$,
as a waiting exposure $P M_T$ for detecting nonzero events of thermal
neutrinos from binary neutron-star mergers with probability 63\%. If the
observation period is taken to be $xP$ for a given value of $M_T$,
detection probability is modified to $1 - e^{-x}$. Here, $E_{\Delta t}$
and $\langle E \rangle$ should be regarded as values averaged over
astrophysical populations of binary neutron stars. The meaning of
factors $f$ is summarized in Table \ref{table:factor}. We note that the
number of mergers during \SI{80}{years} is $\approx \num{2700}$ and that
the expected number of nearby mergers at $\lesssim \SI{3}{Mpc}$ is less
than $0.01$. As we describe in detail in Sec.~\ref{sec:contami},
focusing on $\Delta t_\mathrm{obs} \approx \SI{1}{\second}$ reduces the
length of data from neutrino detectors by a factor of
$\SI{2700}{\second} / \SI{80}{years} \approx \num{e-6}$ compared to a
blind search, dramatically suppressing contamination from other sources.

The period given by Eq.~\eqref{eq:time} is not very short but in the
human time scale for a $\sim \SI{1}{\mega\tonne}$ detector. As a
reference, Hyper-Kamiokande is expected to achieve
\SI{0.37}{\mega\tonne} in the near future, and another Hyper-Kamiokande
is planned to be built in Korea with \SI{0.26}{\mega\tonne}
\cite{hk2016}. One benchmark for the acceptable waiting time is provided
by Galactic supernovae, which are expected to occur once in
30--\SI{100}{years}. Therefore, the detection of thermal neutrinos from
binary neutron stars may be as likely as that from Galactic supernovae,
whereas the number of neutrino events are drastically different. Another
(but related) difference is that binary neutron-star mergers will be
observed steadily by gravitational waves, while a Galactic supernova is
intrinsically rare (see also Ref.~\cite{ando_by2005}). Furthermore, the
prospect for constraining the neutrino mass can be higher for binary
neutron-star mergers than for supernovae due to longer distances as we
will describe in Sec.~\ref{sec:phys_mass}.

In this estimation as well as in Eq.~\eqref{eq:numexp}, we made several
assumptions on astrophysical inputs such as the total energy, typical
energy, spectrum, and merger rate. The most important assumption may be
that remnant massive neutron stars do not collapse before sizable
emission of neutrinos, and this is what we would like to verify or
reject by detecting thermal neutrinos. The total energy is uncertain by
a factor of order unity even within a long-lived remnant scenario and
also depends on the duration of each observation, which we assume to be
$\Delta t_\mathrm{obs} \approx \SI{1}{\second}$. The typical energy is
also uncertain by a factor of $\approx 2$. Spectral deformation is
likely to be a minor correction that can be absorbed in the variation of
$f_E$ (see Table \ref{table:fdinteg}). While the merger rate is highly
uncertain even after the discovery of GW170817, the fiducial value
adopted here is on the conservative side (note also that this value was
denoted as ``realistic'' in Ref.~\cite{ligovirgo2010}). In any case, the
merger rate will be understood in the near future by ongoing
gravitational-wave observations. If some of these parameters conspire,
the waiting time, $P$, could be shortened by a factor of $\gtrsim 5$.

\begin{figure}
 \includegraphics[width=0.95\linewidth]{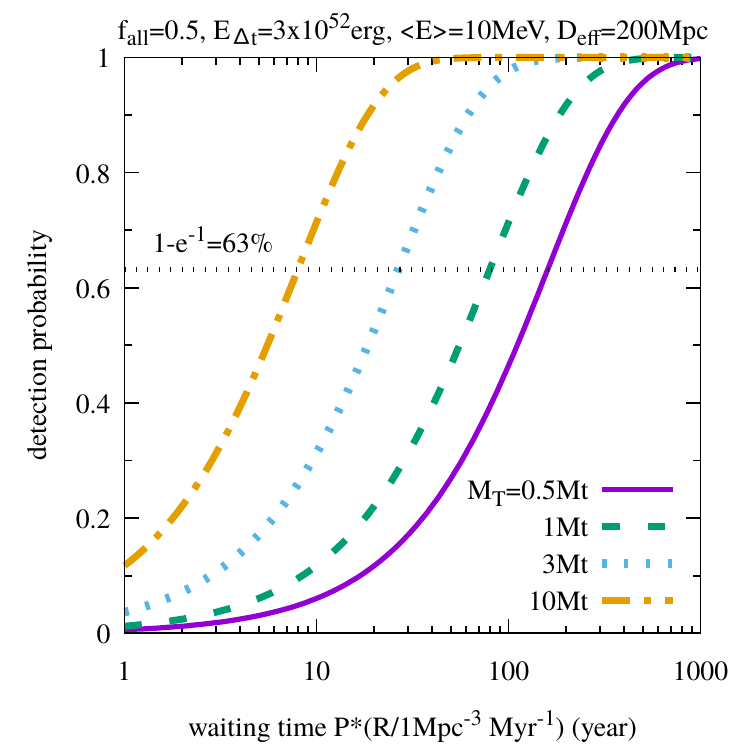} \caption{Detection
 probability of nonzero events of thermal neutrinos as a function of
 time for various detector volumes. The waiting time is normalized to
 our fiducial merger rate of binary neutron stars. Other parameters are
 taken to be our fiducial values shown in Eq.~\eqref{eq:time}. The
 horizontal dotted line indicates $1-e^{-1} = 63\%$.} \label{fig:prob}
\end{figure}

Ultimately, a large effective volume of water- or ice-Cherenkov
detectors is highly desired to increase the likelihood for detecting
thermal neutrinos. Figure \ref{fig:prob} shows the detection probability
of nonzero events of thermal neutrinos as a function of time for various
detector volumes with our fiducial parameters. The effective volume can
be increased not only by a large detector but also by additional
detectors such as Hyper-Kamiokande in Korea. It is conceivable that a
more-than-M-ton detector like Deep-TITAND \cite{kistler_yabs2011} and
MICA \cite{boser_kssv2015} can be constructed within 30--\SI{100}{years}
considered in our strategy. The waiting time can be reduced to less than
\SI{10}{years} with $M_T \gtrsim \SI{8}{\mega\tonne}$ for our fiducial
parameters, and then detections of thermal neutrinos from multiple
binary neutron-star mergers (although one for each) could become
possible within a realistic operating time of telescopes. For example,
Baksan Underground Scintillation Telescope has been running longer than
\SI{30}{years} \cite{baksan2016}. Even for such a large detector, and,
in fact, irrespective of the detector volume, $\Delta t_\mathrm{obs}
\approx \SI{1}{\second}$ has to be chosen to detect thermal neutrinos
from binary neutron-star mergers with high significance as we discuss
below.

\section{Contamination} \label{sec:contami}

\begin{figure}
 \includegraphics[width=0.95\linewidth]{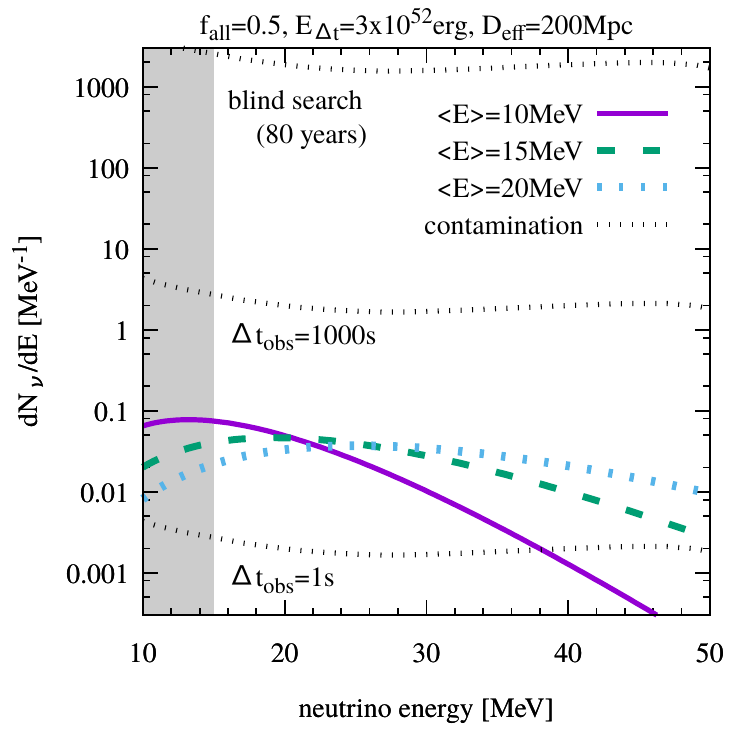} \caption{Spectrum of
 detected thermal neutrinos normalized to a single event between 10 and
 \SI{50}{\mega\eV}, i.e., $\int_{\SI{10}{\mega\eV}}^{\SI{50}{\mega\eV}}
 (d N_\nu /dE) dE = 1$. Purple-solid, green-dashed, and blue-dotted
 curves show the expected spectra at the detector for $\langle E \rangle
 = 10$, 15, and \SI{20}{\mega\eV}, respectively, where we adopt
 Eq.~\eqref{eq:cs} as the cross section. Other parameters are taken to
 be our fiducial values adopted in Eq.~\eqref{eq:time}. Black dotted
 curves show the expected spectrum of contamination during a blind
 search of \SI{80}{years} expected to be required for a single detection
 (top), currently adopted $\Delta t_\mathrm{obs} = \SI{1000}{\second}$
 for all the mergers during this period (middle), and $\Delta
 t_\mathrm{obs} = \SI{1}{\second}$ we proposed in this work
 (bottom). Specifically, we consider invisible muons, atmospheric
 antineutrinos, neutral-current quasielastic scattering, and diffuse
 supernova background as the sources of contamination assuming Gd
 dissolution \cite{hk2011,beacom_vagins2004}. The shaded area on the
 left approximately represents the energy range unavailable due to the
 spallation background, where the precise location of the threshold will
 change.} \label{fig:spec}
\end{figure}

Figure \ref{fig:spec} shows the expected spectrum of thermal neutrinos
from binary neutron-star mergers at water-Cherenkov detectors with the
same level of contamination as planned Hyper-Kamiokande with Gd
dissolution \cite{beacom_vagins2004,hk2011}. To clarify dependence of
the background level on observation strategies, we plot the expected
spectrum of contamination events per single thermal neutrino for (i) a
blind search of \SI{80}{years}, (ii) $\Delta t_\mathrm{obs} =
\SI{1000}{\second}$ for each merger, and (iii) $\Delta t_\mathrm{obs} =
\SI{1}{\second}$ for each merger proposed in this work. Specifically, we
show the sum of decay electrons from invisible muons, atmospheric
antineutrinos, neutral-current quasielastic scattering, and diffuse
supernova neutrino background \cite{hk2011,sekiya2017}. All these
spectra are independent of the detector volume as far as the number of
contamination events scales linearly with the detector volume, except
for the change in the waiting time $P \approx \SI{80}{years}$ for our
fiducial parameters. If the merger rate is changed, the level of
contamination for the blind search scales linearly with the waiting
time, $P$, while the spectra for $\Delta t_\mathrm{obs} =
\SI{1000}{\second}$ and \SI{1}{\second} are unchanged because of the
identical number of mergers.

This figure clearly shows that the chance of detecting thermal neutrinos
arises only when we focus on a short time interval of $\Delta
t_\mathrm{obs} \approx \SI{1}{\second}$ right after the
merger. Otherwise, for example with the currently adopted $\Delta
t_\mathrm{obs} = \SI{1000}{\second}$ \cite{kamland2016,sk2016}, thermal
neutrinos from binary neutron-star mergers are heavily obscured by
contamination from other sources of neutrinos. In the following, we
discuss the expected level of contamination more quantitatively.

\subsection{Quantitative assessment} \label{sec:contami_quant}

\begin{table*}
 \caption{Background and the expected number of events for
 \SI{2700}{Mt.s} relevant for detecting a single thermal neutrino from
 binary neutron-star mergers (but note that the waiting exposure depends
 on the energy window via $f_E$). All the data are taken from
 Ref.~\cite{hk2011}, except for neutral-current quasielastic scattering
 taken from Ref.~\cite{sekiya2017} by assuming that Gd does not change
 the level of this background. We do not include solar and reactor
 neutrinos severe at $E \lesssim \SI{10}{\mega\eV}$.}
 \begin{tabular}{cccc} \hline
  & Without Gd (20--\SI{30}{MeV}) & With Gd (10--\SI{50}{MeV})\\
  \hline \hline
  Decay electron from invisible muons & 0.03 & 0.03 \\
  Atmospheric antineutrino & 0.003 & 0.003 \\
  Neutral-current quasielastic scattering & $< \num{e-3}$ & 0.003 \\
  Diffuse supernova neutrino background & 0.004 & 0.01 \\
  \hline
 \end{tabular}
 \label{table:contami}
\end{table*}

We examine how severe contamination from other sources of neutrinos is
to detect thermal neutrinos from binary neutron-star mergers focusing on
Hyper-Kamiokande based on Ref.~\cite{hk2011}. Characteristics of target
neutrinos are very similar to those from supernova explosions, and thus
backgrounds are basically the same as those encountered in searches of
the diffuse supernova neutrino background \cite{beacom2010}. This fact
implies that Gd dissolution will significantly increase the prospect for
detecting thermal electron antineutrinos from binary neutron-star
mergers via tagging neutrons from inverse $\beta$ decay
\cite{beacom_vagins2004}. Important numbers are summarized in Table
\ref{table:contami} and the final paragraph of this subsection.

Without Gd, we have to cope not only with electron antineutrinos that
induce inverse $\beta$ decay but also with various sources of Cherenkov
radiation. On one hand, the lower energy threshold, $E_\mathrm{min}$,
will be required to be $\gtrsim \SI{20}{\mega\eV}$ to avoid spallation
products and solar neutrinos. As a reference, the number of solar
neutrino events in 9.0--\SI{9.5}{\mega\eV} is reported to be \num{1350}
for \SI{0.09}{Mt.years} in Super-Kamiokande \cite{hk2011}. This
corresponds to $\approx 1.3$ events for \SI{2700}{Mt.s} relevant for
detecting a single thermal neutrino, and the rate for spallation
products is higher at least by a factor of 5 than this. Even though
solar neutrinos become much weaker at higher energy (see also
discussions in Ref.~\cite{ando_st2003}), spallation products will serve
as severe contaminants for Hyper-Kamiokande at a shallow site (but see
Refs.~\cite{li_beacom2014,li_beacom2015,li_beacom2015-2} for possible
order-of-magnitude reduction informed by shower physics). On the other
hand, the higher threshold $E_\mathrm{max}$ has to be chosen to avoid
decay electrons from invisible muons. The rate of events from invisible
muons (and atmospheric electron antineutrinos) is expected to be
$\approx 220$ for 20--\SI{30}{\mega\eV} and \SI{0.56}{Mt.years}
\cite{hk2011}. Thus, it will produce only $\approx 0.03$ event for
\SI{2700}{Mt.s}, while the number will increase to $\approx 0.08$ and
0.15 with $E_\mathrm{max} = \SI{40}{\mega\eV}$ and \SI{50}{\mega\eV},
respectively. Therefore, a reasonable choice may be $E_\mathrm{min}
\approx \SI{20}{\mega\eV}$ and $E_\mathrm{max} \approx
\SI{30}{\mega\eV}$, which results in $f_E \approx 0.2$--0.33 (see Table
\ref{table:fdinteg}). This could be acceptable but is not very
satisfactory. In particular, the number of mergers required for
detecting a thermal neutrino is proportional to $f_E^{-1}$, and thus
contamination events increase accordingly.

If Gd is dissolved successfully, the situation improves
substantially. First, spallation backgrounds as well as solar neutrinos
can be suppressed via neutron tagging. The lower threshold may be
reduced to $E_\mathrm{min} \lesssim \SI{10}{\mega\eV}$ limited by
electron antineutrinos from nuclear reactors. A caveat comes from
$\beta^- n$ decay of isotopes such as ${}^9$Li and ${}^8$He, which
mimics the inverse $\beta$ decay. Still, it is suggested that these
isotopes could be removed by identifying preceding neutrons produced
during the spallation \cite{li_beacom2014}, and here we optimistically
assume that these isotopes can also be efficiently removed. Because
neutrinos with $E < \SI{10}{\mega\eV}$ are minor (see Table
\ref{table:fdinteg}), it will be sufficient if we could reduce
$E_\mathrm{min}$ to \SI{10}{MeV}. Next, Gd also reduces the events from
invisible muons at high energy by a factor of $\sim 5$, and this allows
us to choose $E_\mathrm{max} = 40$--\SI{50}{\mega\eV} with keeping the
rate of contamination to be $\lesssim 0.02$--0.04 for \SI{2700}{Mt.s}
taking the reduction of selection efficiency, $f_\mathrm{se}$, into
account. These threshold values give us $f_E \approx 0.8$.

Recently, quasielastic scattering of neutrinos by oxygen nuclei via
neutral-current interactions has been recognized as a significant source
of contamination at low energy (see, e.g.,
Ref.~\cite{priya_lunardini2017}). Because this interaction could mimic
inverse $\beta$ decay via neutron ejection, Gd cannot be used to
suppress this background in a straightforward manner. This contamination
could dominate invisible muons at $\lesssim \SI{15}{\mega\eV}$, and the
number of events is estimated to be $\approx 0.003$ for
10--\SI{50}{\mega\eV} and \SI{2700}{Mt.s} according to
Ref.~\cite{sekiya2017}. We note that this contamination has not yet been
studied extensively, and further reduction is discussed for detecting
the diffuse supernova neutrino background \cite{priya_lunardini2017}.

One difference from the search of the diffuse supernova neutrino
background is that these neutrinos themselves serve as contamination in
our search. The event rate is estimated to be $\approx 83$ in
10--\SI{30}{\mega\eV} for \SI{0.56}{Mt.years} with Gd dissolution
\cite{hk2011}, which corresponds to $\approx 0.01$ for
\SI{2700}{\second} with a \SI{1}{\mega\tonne} detector, and the
contribution from $E > \SI{30}{\mega\eV}$ is minor. While the
uncertainty is large, it is not very likely that the realistic diffuse
background is very intense taking present nondetection into account.

The expected number of contamination events per single thermal neutrino
from binary neutron-star mergers $r$ and the required energy window are
summarized as follows (see also Table \ref{table:contami}). If Gd is not
dissolved, we have to choose 20--\SI{30}{\mega\eV} to avoid spallation
products at low energy and decay electrons from invisible muons at high
energy. Taking the increase of required exposure for a single detection
by a factor of 2--3, this will result in $r \sim 0.05$--0.1. If Gd is
dissolved, we may be able to achieve $r \approx 0.03$--0.05 adopting
10--40 or \SI{50}{\mega\eV}. The lower threshold is determined by
reactor neutrinos, and the higher threshold is determined by invisible
muons now suppressed by a factor of $\sim 5$.

\subsection{Toward high significance} \label{sec:contami_sig}

The contamination event $r \sim 0.03$ for one detection of thermal
neutrinos in \SI{80}{years} is not hopeless but not very
comfortable. Straightforward improvement comes from a large detector
that will enable us to detect multiple thermal neutrinos. Here, we would
like to discuss other directions to reduce the contamination further.

For this purpose, dependence of the number of contamination events on
various parameters should be examined. Generally, strong neutrino
emission per merger reduces the required number of mergers and increase
the significance of detections. Thus, large values of $f_\mathrm{all}$,
$E_{\Delta t}$, and $\langle E \rangle$ reduce the number of
contamination events. At the same time, a large number of mergers and a
long observing time window will increase the number of contamination
events. Specifically, the expected number of contamination events per
single thermal neutrino from binary neutron-star mergers is given by
\begin{equation}
 r \propto \frac{\Delta t_\mathrm{obs} D_\mathrm{eff}^2}{f_\mathrm{all}
  E_{\Delta t} \langle E \rangle} .
\end{equation}
Here, dependence on $D_\mathrm{eff}$ is given by competition between the
volume $\propto D_\mathrm{eff}^3$ and the period required for detecting
a single neutrino $P \propto D_\mathrm{eff}^{-1}$ [see
Eq.~\eqref{eq:time}].

One parameter we can actively choose is $\Delta t_\mathrm{obs}$, which
also affects $E_{\Delta t}$. Because the neutrino luminosity is higher
in the earlier epoch, focusing on a short time window after the merger
is advantageous for increasing the significance. For example, we may be
able to choose $\Delta t_\mathrm{obs} = \SI{0.1}{\second}$ while
reducing $E_{\Delta t}$ only by a factor of $\approx 3$. This gives us
$r \approx 0.01$ with an obvious price of increasing the waiting time by
the same factor. Accurate numerical simulations of neutrino emission
will be helpful to determine an optimal time interval, $\Delta
t_\mathrm{obs}$, and energy thresholds, $\{ E_\mathrm{min} ,
E_\mathrm{max} \}$, for detecting thermal neutrinos with high
significance.

Another parameter we can actively choose is $D_\mathrm{eff}$, the
distance to which we try to observe thermal neutrinos from binary
neutron-star mergers, or equivalently the threshold signal-to-noise
ratio for gravitational-wave detections. If we discard distant mergers
with weak neutrino emission, the average fluence of neutrinos per merger
increases. Again, the price is the increase of the waiting time,
$P$. Because $r P^2$ is approximately independent of $D_\mathrm{eff}$,
the number of contamination events can be suppressed relatively
efficiently by restricting the range of $D_\mathrm{eff}$ with only a
modest increase of the waiting time. Specifically, the waiting time
increases only by $\sqrt{3} \approx 1.7$ when $r$ is reduced from
$\approx 0.03$ to $\approx 0.01$. The reason for this is that we
selectively keep nearby binary neutron-star mergers with large fluences
of neutrinos. Therefore, we expect that observing binary neutron-star
mergers within $D_\mathrm{eff} \lesssim \SI{120}{Mpc}$ with $\Delta
t_\mathrm{obs} \approx \SI{1}{\second}$ may be close to the optimal
strategy. This consideration on $D_\mathrm{eff}$ immediately means that
high-sensitivity gravitational-wave detectors such as the Einstein
Telescope \cite{et2012} and Cosmic Explorer \cite{ce2017} will not
necessarily be helpful to detect thermal neutrinos, because the
significance can be kept high only when we focus on nearby mergers.

The level of contamination will be independent of the merger rate,
$\mathcal{R}$, because it is irrelevant to the neutrino energy emitted
during $\Delta t_\mathrm{obs}$. While the merger rate critically affects
the waiting time, $P$, its uncertainty does not affect the significance
of neutrino detections achieved with our strategy. We also do not expect
that the value of $M_T$ changes the significance of each detection of a
thermal neutrino.

\section{Physics implication} \label{sec:phys}

Even though we may be able to detect only a single thermal neutrino, it
offers a unique opportunity to extract various information. In this
section, we describe its possible implication to physics and
astrophysics of neutrinos.

\subsection{Energy scale of the neutrino emission}
\label{sec:phys_energy}

\begin{figure}
 \includegraphics[width=0.95\linewidth]{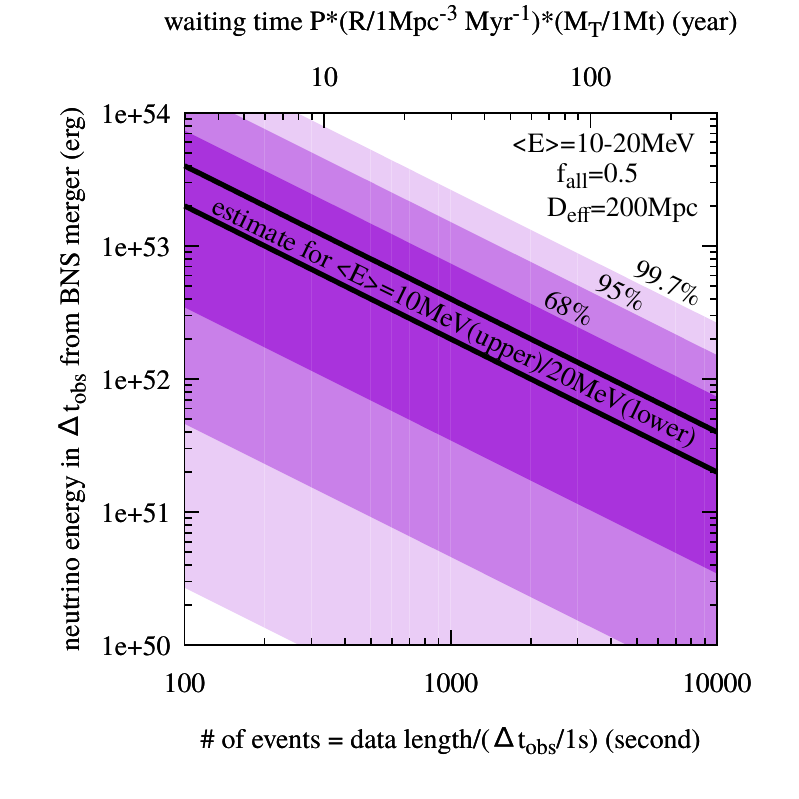} \caption{Confidence
 interval of the neutrino energy emitted in $\Delta t_\mathrm{obs}$ as a
 function of the number of binary neutron-star mergers observed until
 the first detection. The upper and lower solid lines denote the
 estimates under the assumption of $\langle E \rangle = \SI{10}{MeV}$
 and \SI{20}{MeV}, respectively.} \label{fig:energy}
\end{figure}

We can infer the energy scale of the neutrino emission by counting the
number of mergers that we collect to detect a single thermal
neutrino. Figure \ref{fig:energy} shows the confidence interval of the
neutrino energy that typical binary neutron-star mergers emit during
$\Delta t_\mathrm{obs}$, namely $E_{\Delta t}$. We may be able to narrow
down the energy scale to about an order of magnitude with 68\%
confidence even accounting for a factor of $\approx 2$ uncertainty in
the typical energy, $\langle E \rangle$.

While this constraint is crude, this gives us a unique opportunity to
characterize the emission of thermal neutrinos from binary neutron-star
mergers in a quantitative manner. In particular, this constraint is the
most direct route to infer the nature of the merger remnant, e.g.,
formation of a long-lived neutron star. Emission with $\gtrsim
\SI{e52}{erg}$ supports formation of massive neutron stars as a typical
remnant of binary neutron stars rather than a prompt collapse, whereas
formation of a massive black hole accretion disk with $\gtrsim 0.1
M_\odot$ is not strictly excluded. The upper limit on $E_{\Delta t}$
will reject extraordinary emission of thermal neutrinos. If $E_{\Delta
t}$ is smaller than, say \SI{e52}{erg}, it would indicate that either
the neutrino emission is relatively weak or the merger remnant rapidly
collapses to a black hole. The former will be true if the neutron-star
equation of state is very stiff so that the collision does not increase
the temperature very much or if the viscosity does not play an important
role to heat up the remnant material. The latter should indicate that
the maximum mass of the neutron star is not very large, say very close
to $2 M_\odot$.

\subsection{Constraining the neutrino mass} \label{sec:phys_mass}

Once we detect a thermal neutrino by monitoring multiple mergers, we can
identify the progenitor gravitational-wave source from time coincidence
in a straightforward manner. One possible concern is that neutrinos may
be delayed substantially from gravitational waves due to their finite
masses so that the time coincidence becomes loose or completely lost
(say, time delay longer than a day). Specifically, the velocity of
neutrinos with the momentum $p$ is given by
\begin{equation}
 \frac{v}{c} \approx 1 - \frac{m_\nu^2 c^2}{2 p^2}
\end{equation}
for a neutrino mass $m_\nu \ll p/c \approx E/c^2$. By contrast, because
the dispersion relation of gravitational waves, or gravitons, is
securely constrained to be that of massless particles with an upper
limit of $\approx \SI{e-22}{\eV} / c^2$ on the corresponding graviton
mass \cite{ligovirgo2017}, we can safely assume that gravitational waves
propagate with the speed of light.\footnote{It should be cautioned that
frequency-independent modification of the speed of gravity cannot be
constrained meaningfully from current gravitational-wave
observations. We neglect this anomalous case. The constraint obtained by
comparing GW170817 and GRB 170817A is not sufficient for our purpose
\cite{ligovirgogamma2017}.} Thus, the expected time delay of neutrinos
relative to gravitational waves is written as
\begin{align}
 \Delta t_d & = \left( 1 - \frac{v}{c} \right) \frac{D}{c} \\
 & \approx \SI{0.51}{\second} \left( \frac{D}{\SI{100}{Mpc}} \right)
 \left( \frac{m_\nu c^2}{\SI{0.1}{\eV}} \right)^2 \left(
 \frac{E}{\SI{10}{\mega\eV}} \right)^{-2} .
\end{align}

This expression implies that the time delay can become problematic only
for low-energy neutrinos from a very distant merger. Hereafter, the
three mass eigenvalues are denoted as $m_1$, $m_2$, and $m_3$. Because
$m_1$ $( \approx m_2 )$ should be smaller than $\SI{0.1}{\eV} / c^2$
taking the squared mass difference $| \Delta m_{31}^2 | \approx | \Delta
m_{32}^2 | \approx \SI{2.5e-3}{\square\eV} / c^4 = ( \SI{0.05}{\eV} /
c^2 )^2$ \cite{capozzi_dlmmp2017} and an upper limit on the sum of three
mass eigenvalues $\sum_{i=1,2,3} m_i \lesssim \SI{0.2}{\eV} / c^2$
inferred from the \textit{Planck} measurement combined with baryon
acoustic oscillation measurements \cite{planck2016}, the realistic time
delay should be much smaller than the assumed duration of each analysis,
$\Delta t_\mathrm{obs} \approx \SI{1}{\second}$, particularly for normal
hierarchy. Even if the hierarchy is inverted, the dominant part with
$m_1 \approx m_2$ is marginally able to produce $\Delta t_d \gtrsim
\SI{1}{\second}$ with a ``worst'' combination of parameters. This fact
means that the time delay will not substantially degrade the performance
of our strategy for neutrino detections. Accordingly, we do not have to
worry seriously about the reduction of $f_\mathrm{osc}$ due to
broadening in time of neutrino light curves and corresponding decrease
of the flux caused by the mass differences.

Conversely, if we could detect a neutrino and measure its time delay
$\Delta t_d$ relative to the merger, or relative time of flight, we can
impose an upper limit on the absolute neutrino mass of the lightest
eigenstate\footnote{Because $m_3$ is coupled only very weakly to the
electron-type neutrino \cite{capozzi_dlmmp2017}, we may be able to
constrain the mass of $m_1 \approx m_2$, which is not the lightest for
inverted hierarchy. We try to be conservative here, because the neutrino
oscillation is the source region is not fully understood.} from the
condition that the mass should not produce time delay longer than
$\Delta t_d$. Quantitatively, we immediately derive
\begin{align}
 m_\nu c^2 & \lesssim \sqrt{\frac{2 c \Delta t_d}{D}} E \\
 & \approx \SI{44}{\milli\eV} \left( \frac{\Delta
 t_d}{\SI{0.1}{\second}} \right)^{1/2} \left( \frac{D}{\SI{100}{Mpc}}
 \right)^{-1/2} \left( \frac{E}{\SI{10}{\mega\eV}} \right) .
 \label{eq:mconstr}
\end{align}
Here, we adopt $\Delta t_d = \SI{0.1}{\second}$ somewhat optimistically,
and we believe that we have a good chance to obtain this value because
of the higher luminosity in the earlier epoch
\cite{sekiguchi_kks2011,fujibayashi_sks2017}. In principle, $\Delta t_d
\approx \SI{1}{\milli\second}$ can be achieved, where the limitation
comes from the timing accuracy of gravitational-wave
detectors. Hyper-Kamiokande will determine the arrival time of neutrinos
much more accurately than \SI{1}{\milli\second}.

The measurement error of the distance, $D$, could degrade the constraint
significantly.  Gravitational-wave detectors will determine the distance
within an error of $\approx 50\%$ \cite{ligovirgo2016-5}, and thus the
constraint on the neutrino mass will be loosened by $\approx 25\%$. The
accuracy of the distance measurement can be improved by an order of
magnitude and thus become negligible if we detect electromagnetic
counterparts, which can be searched for after prompt identification of a
coincident neutrino.

Equation \eqref{eq:mconstr} suggests that we might be able to constrain
the absolute neutrino mass of the lightest eigenstate to $\lesssim
O(10)\,\si{\milli\eV} / c^2$ by detecting a thermal neutrino from binary
neutron-star mergers. It would be worthwhile to compare this value with
other proposals for constraints. \emph{(i) Supernova.} This limit is
tighter by an order of magnitude than \si{\eV}-scale constraints
envisioned for supernova observations both without gravitational waves
\cite{totani1998,beacom_bm2000} and with gravitational waves
\cite{arnaud_bbcdhp2002}. The primary reason of this improvement is that
binary neutron stars merge at cosmological distances of $\gtrsim
\SI{100}{Mpc}$, which should be compared with the length scale of our
Galaxy, \SI{10}{kpc}. \emph{(ii) Direct measurements.}  The KATRIN
experiment is now planning to measure directly the effective mass of
electron neutrinos down to $m_{\nu_e,\mathrm{eff}} \approx \SI{0.2}{\eV}
/ c^2$ via the $\beta$ decay of the tritium \cite{katrin2013}, although
it is not fair to compare future observations on a time scale of
$\approx 30$--\SI{100}{years} considered in this study with ongoing
experiments. Double $\beta$ decay experiments can also constrain the
mass of neutrinos to sub-\si{\eV} if they are Majorana particles, but
this limit does not apply to Dirac neutrinos (see
Ref.~\cite{avignone_ee2008} for reviews). \emph{(iii) Cosmology.}  Our
constraint is comparable to current cosmological constraints
\cite{palanquedelabrouille_etal2015,planck2016} (see also
Ref.~\cite{giusarma_gmvhf2016,vagnozzi_gmfghl2017}). Taking the
potential uncertainty of cosmological models into account (see, e.g.,
Ref.~\cite{riess_etal2016}), the independent constraint from the
relative time-of-flight will be invaluable.

\section{Summary} \label{sec:summary}

We presented a long-term strategy to detect thermal neutrinos emitted
from the remnant of binary neutron-star mergers with a future M-ton
water-Cherenkov detector such as Hyper-Kamiokande
\cite{hk2011,hk2016}. Although the detection from a single merger is not
expected and the diffuse neutron-star-merger neutrino background will be
hidden by other neutrinos, monitoring multiple mergers only for $\Delta
t_\mathrm{obs} \approx \SI{1}{\second}$ each by using timing information
from gravitational-wave detectors could give us a chance of detection
with a human time-scale operation of $\approx
\SI{80}{Mt.years}$. Contamination from other sources of neutrinos may be
reduced to $\approx 0.03$ with Gd dissolution. We may be able to reduce
the contamination further to $\approx 0.01$ with an increase of the
waiting time by only a factor of $\approx 1.7$ by focusing only on
slightly nearby mergers. Ultimately, the chance of detections can be
increased by a large effective volume of neutrino detectors, and
possible more-than-M-ton class detectors such as Deep-TITAND
\cite{kistler_yabs2011} and MICA \cite{boser_kssv2015} will enable us to
detect thermal neutrinos from multiple binary neutron-star mergers.

The direct detection will qualitatively confirm the formation of a hot
remnant after the merger of binary neutron stars and verify current
theoretical pictures. Moreover, the energy scale of the neutrino
emission can be constrained from the number of mergers that we collect
to detect a single neutrino, and the formation of remnant massive
neutron stars could be strongly supported or disfavored. Because
distances to binary neutron-star mergers are expected to be cosmological
($\gtrsim \SI{100}{Mpc}$), we could obtain meaningful upper limits,
$\lesssim O(10)\,\si{\milli\eV} / c^2$, on the absolute neutrino mass of
the \emph{lightest} eigenstate from the time delay relative to
gravitational waves, which are now securely considered to propagate with
the speed of light.

\begin{acknowledgments}
 We are deeply indebted to John F. Beacom for careful reading of the
 manuscript and helpful comments. We also thank Kazunori Kohri, Yusuke
 Koshio, and Takaaki Yokozawa for valuable discussions, and Shin'ichiro
 Ando, Kunihito Ioka, Kohji Ishidoshiro, Kenta Kiuchi, Marek Kowalski,
 and Tomonori Totani for useful comments on the earlier version of the
 manuscript. We are also grateful to Takashi Nakamura for a comment on
 the preprint version and to an anonymous referee for informing us about
 neutral-current quasielastic scattering. This work is supported by the
 Japanese Society for the Promotion of Science (JSPS) KAKENHI
 Grant-in-Aid for Scientific Research (No.~JP16H06342, No.~JP17H01131,
 No.~JP17K14248).
\end{acknowledgments}

\appendix

\section{Computation of $f_\Omega$} \label{sec:antenna}

The strain received by a gravitational-wave detector like Advanced LIGO
is given by (see, e.g., Ref.~\cite{sathyaprakash_schutz2009})
\begin{equation}
 h ( \theta , \varphi , \imath , \psi ) = F_+ ( \theta , \varphi ,
  \psi ) h_+ ( \imath ) + F_\times ( \theta , \varphi , \psi ) h_\times
  ( \imath ) .
\end{equation}
The antenna pattern functions $\{ F_+ , F_\times \}$ depend on the
position of the source on the sky $( \theta , \varphi )$ and the
so-called polarization angle $\psi$ that dictates the orientation of the
source in the sky plane with respect to the detector as
\begin{align}
 F_+ & = \frac{1}{2} ( 1 + \cos^2 \theta ) \cos 2 \varphi \cos 2 \psi -
 \cos \theta \sin 2 \varphi \sin 2 \psi , \\
 F_\times & = \frac{1}{2} ( 1 + \cos^2 \theta ) \cos 2 \varphi \sin 2
 \psi + \cos \theta \sin 2 \varphi \cos 2 \psi .
\end{align}
For a quadrupolar gravitational-wave source with the inclination angle
$\imath$, we have
\begin{equation}
 h_+ = h_0 \frac{1 + \cos^2 \imath}{2} \cos ( \omega t ) \; , \;
  h_\times = h_0 \cos \imath \sin ( \omega t ) ,
\end{equation}
where $h_0$ and $\omega$ are the amplitude and frequency of
gravitational waves, respectively. To separate geometrical parameters
from intrinsic properties of the source, it is useful to define
\begin{equation}
 w \equiv \sqrt{ \frac{( 1 + \cos^2 \imath )^2}{4} F_+^2 ( \theta ,
  \varphi , \psi ) + \cos^2 \imath F_\times^2 ( \theta , \varphi , \psi
  )} ,
\end{equation}
where $w \le 1$. The inequality is saturated for face-on binaries
($\imath = 0$ or $\pi$) along the normal direction to the detector plane
($\theta = 0$ or $\pi$). The horizon distance $D_H$ is defined as the
maximal distance at which the signal is detectable with a threshold
signal-to-noise ratio and is realized for $w=1$.

The detectable volume is given in terms of the horizon distance, $D_H$,
by averaging over the binary orientation $( \imath , \psi )$ and
integrating over the sky position $( \theta , \varphi )$ as
\begin{equation}
 V \equiv \frac{1}{4\pi} \int_{\imath \psi} \int_{\theta \varphi}
  \int_0^{D_H w} r^2 dr d\Omega_{\theta \varphi} d \Omega_{\imath \psi}
  ,
\end{equation}
and the effective range of the detector is related to the detectable
volume by $V = 4 \pi D_\mathrm{eff}^3 / 3$. The integral can be
evaluated numerically as
\begin{equation}
 \frac{1}{( 4 \pi )^2} \int_{\imath \psi} \int_{\theta \varphi} w^3
  d\Omega_{\theta \varphi} d\Omega_{\imath \psi} \approx
  \frac{1}{(2.26)^3} ,
\end{equation}
and this shows that $D_\mathrm{eff} \approx D_H / 2.26$. This should be
compared with $D_\mathrm{eff} = D_H$ for a hypothetical case with $w=1$
in all the directions and orientations.

The neutrino flux and fluence are proportional to $D^{-2}$, and the
average fluence of neutrinos from all the mergers detectable by
gravitational waves is given by
\begin{equation}
 S_\mathrm{ave} = \frac{E_{\Delta t}}{4\pi V} \times \frac{1}{4\pi}
  \int_{\imath \psi} \int_{\theta \varphi} \int_0^{D_H w} dr
  d\Omega_{\theta \phi} d\Omega_{\imath \varphi} .
\end{equation}
For a hypothetical case with $w=1$ in all the directions and
orientations, this gives us $S_\mathrm{ave} = E_{\Delta t} D_H / V =
E_{\Delta t} D_\mathrm{eff} / V$. For the realistic antenna pattern, we
numerically obtain
\begin{equation}
 \frac{1}{( 4 \pi )^2} \int_{\imath \psi} \int_{\theta \varphi}
  w d\Omega_{\theta \varphi} d\Omega_{\imath \psi} \approx 0.352 ,
\end{equation}
and the average fluence of neutrinos is found to be reduced by a factor
of $0.352 \times 2.26 \approx 0.797$ compared to the hypothetical
case for a given value of $D_\mathrm{eff}$.

%\bibliography{paper}
%
\end{document}